\documentclass{emulateapj}
\shorttitle{The Reverse Shock of SN1987a}
\shortauthors{Smith et al.}
\begin{document}

\title{THE REVERSE SHOCK OF SNR1987A AT 18 YEARS AFTER OUTBURST\altaffilmark{1}}

\author{Nathan Smith\altaffilmark{2,3}, Svetozar A.\
Zhekov\altaffilmark{4,8}, Kevin Heng\altaffilmark{4}, Richard
McCray\altaffilmark{4}, Jon A.\ Morse\altaffilmark{5,6}, and Mike
Gladders\altaffilmark{7}}

\altaffiltext{1}{Based on observations made at the Clay Telescope of
  the Magellan Observatory, a joint facility of the Carnegie
  Observatories, Harvard University, the Massachusetts Institute of
  Technology, the University of Arizona, and the University of
  Michigan.}

\altaffiltext{2}{Center for Astrophysics and Space Astronomy, University of
Colorado, 389 UCB, Boulder, CO 80309}

\altaffiltext{3}{Hubble Fellow; nathans@casa.colorado.edu}

\altaffiltext{4}{JILA, University of Colorado, 440 UCB, Boulder, CO
80309}

\altaffiltext{5}{Department of Physics and Astronomy, Arizona State
University, Box 871504, Tempe, AZ 85287-1504}

\altaffiltext{6}{Present address: Observational Cosmology Laboratory,
  Code 665, Goddard Space Flight Center, Greenbelt, MD 20771}

\altaffiltext{7}{Carnegie Observatories, Pasadena, CA 91101}

\altaffiltext{8}{On leave from the Space Research Institute, Sofia, Bulgaria}

\begin{abstract}

We use low-dispersion spectra obtained at the Magellan Observatory to
study the broad H$\alpha$ emission from the reverse shock of the
infant supernova remnant SNR1987A.  These spectra demonstrate that the
spatio-kinematic structure of the reverse shock can be distinguished
from that of the circumstellar ring and hotspots, even at ground-based
spatial resolution.  We measure a total dereddened H$\alpha$ flux of
1.99($\pm$0.22)$\times$10$^{-13}$ ergs s$^{-1}$ cm$^{-2}$ at an epoch
18.00 years after outburst.  At 50 kpc, the total reverse shock
luminosity in H$\alpha$ is roughly 15~$L_{\odot}$, which implies a
total flux of neutral hydrogen atoms across the reverse shock of
8.9$\times$10$^{46}$ s$^{-1}$, or roughly 2.3$\times$10$^{-3}$
$M_{\odot}$ yr$^{-1}$.  This represents an increase by a factor $\sim
4$ since 1997.  Lyman continuum radiation from gas shocked by the
forward blast wave can ionize neutral hydrogen atoms in the supernova
debris before they reach the reverse shock.  If the inward flux of
ionizing photons exceeds the flux of hydrogen atoms approaching the
reverse shock, this pre-ionization will shut off the broad Ly$\alpha$
and H$\alpha$ emission. The observed X-ray emission of SNR1987A
implies that the ratio of ionizing flux to hydrogen atom flux across
the reverse shock is presently at least 0.04. The X-ray emission is
increasing much faster than the flux of atoms, and if these trends
continue, we estimate that the broad Ly$\alpha$ and H$\alpha$ emission
will vanish in $\la$7 years.

\end{abstract}


\section{INTRODUCTION}

The collision between the ejecta of SN1987A and its circumstellar ring
is now in full bloom, signaling the birth of the supernova remnant
SNR1987A.  This interaction was predicted (Luo \& McCray 1991; Luo et
al.\ 1994; Chevalier \& Dwarkadas 1995; Borkowski et al.\ 1997)
shortly after the discovery of the circumstellar ring. It began about
a decade after the outburst with the discovery of the first of many
``hot spots'' (Sonneborn et al.\ 1998; Michael et al.\ 2000; Pun et
al.\ 2002).  These hot spots are thought to result when the forward
blast wave encounters and transmits radiative shocks into dense
protrusions or ``fingers'' pointing to the interior of the ring.
Since then, hot spots have encircled the entire ring (Sugerman et al.\
2002).

Behind the blast wave, the expanding supernova debris are decelerated
by a reverse shock (e.g., Chevalier 1982). This non-radiative shock is
seen as very broad, high-velocity Ly$\alpha$ and H$\alpha$ emission
features in Space Telescope Imaging Spectrograph (STIS) data (Michael
et al.\ 2003, 1998; Sonneborn et al.\ 1998). This emission, which was
predicted by Borkowski et al.\ (1997), results from the collisional
excitation of neutral H atoms from the supernova debris crossing the
shock front.  Using a series of long-slit STIS spectra, Michael et
al.\ (1998) mapped the geometry of the reverse shock, finding it to
reside within roughly $\pm$30\arcdeg\ of the equator.  New
ground-based observations reported here, supplemented by additional
observations with STIS (Sonneborn et al.\ 1998; Heng et al., in
preparation), show that the broad Ly$\alpha$ and H$\alpha$ reverse
shock emission has increased by a factor $\sim$4 since 1997.
Meanwhile, the X-ray emission from SNR1987A has brightened at a
rapidly accelerating rate (Park et al.\ 2005).  The blast wave
interaction with the hot spots now dominates the X-ray emission
(Zhekov et al.\ 2005).

In this Letter we present new spectra from the Magellan Observatory,
which demonstrate that we can continue to study the emission from this
reverse shock with ground-based telescopes, despite the recent demise
of STIS.  We also discuss the possibility that the shocked gas will
produce sufficient ionizing luminosity to photoionize hydrogen atoms
in the supernova debris before they reach the reverse shock, and
thereby suppress the broad Ly$\alpha$ and H$\alpha$ emission.

\section{MAGELLAN OBSERVATIONS}

We observed SN1987A during the commissioning run of the Low Dispersion
Survey Spectrograph-3 (LDSS3) mounted on the Clay Telescope of the
Magellan Observatory on the evening of 2005 Feb 24 (Feb 25 UT), almost
exactly 18 years after the supernova was first discovered.  LDSS3 has
an STA0500A 4064$\times$4064 CCD detector.  At red wavelengths it has
a pixel scale of 0$\farcs$189$\times$1.124 \AA~and an effective
spectral resolution of $\sim$5 \AA \ (R=1300) with a 0$\farcs$8
($\sim$4.5 pixel) slit width.  During the observations the weather was
clear and the seeing was $\sim$0$\farcs$8.  The slit aperture was
centered on the supernova and oriented at P.A.=--10$\arcdeg$, aligned
with the minor axis of the equatorial ring.  Wavelengths and
velocities are measured with respect to the rest wavelengths of the
narrow nebular emission along the slit.  Airglow lines and emission
from the surrounding H~{\sc ii} region were subtracted by fitting the
emission from the background sky along the slit.

The region of the long-slit spectrum around H$\alpha$ resulting from a
total exposure time of 900s is shown in Figure 1$a$.  In order to
flux-calibrate the LDSS3 spectrum in February 2005, we interpolated
between the flux measured in 2004 December and 2005 May in {\it
HST}/ACS images.  We used images in the F658N filter, which includes
narrow H$\alpha$ and [N~{\sc ii}] emission from the circumstellar ring
and hot spots. We extracted the flux over the same spatial region
centered on the northern half of the ring in both the images and the
LDSS spectrum, and summed the flux over the F658N filter bandpass
($\Delta\lambda\simeq$50 \AA) in the spectrum.  Tracings of the
resulting flux-calibrated spectrum are shown in Figure 1$b$, where we
display the total flux within $\pm$1\arcsec, as well as separate
extractions for the northern (blueshifted) and southern (redshifted)
sides of the nebula.

In order to isolate the very broad emission from the reverse shock, we
interpolated across the narrow nebular emission lines from the ring
and hot spots (Figure 1$c$).  We also removed the emission hump due to
rapidly fading H$\alpha$ emission from the central remnant by
subtracting a Gaussian profile (had we included this emission, the
total flux would have been 6.5\% higher).  Interpolating across
He~{\sc i} $\lambda$6680 and the red [S~{\sc ii}] doublet was somewhat
subjective; we relied on what appeared to be reverse shock emission at
$\sim$6700 \AA\ between the two sets of lines.  This resulted in a red
bump in the reverse shock line profile.  This is probably the best
representation of the true reverse shock emission, and is used for the
estimate below.  A very conservative estimate made by interpolating
linearly from 6660 to 6760 \AA\ (the dotted or blue line in Figure
1$c$) would have resulted in a total flux about 6\% less than the
value we quote below.

We measure a total H$\alpha$ flux for the broad reverse shock
component of 1.37($\pm$0.15)$\times$10$^{-13}$ ergs s$^{-1}$ cm$^{-2}$
on 2005 February 25.  This corresponds to the black tracing in Figure
1$c$, integrated from 6200 to 6860 \AA, and is continuum-subtracted
using the continuum level shown by the straight dashed line in Figure
1$c$.  This estimate includes an adjustment of +15\% for those parts
of the reverse shock on the east and west edges of the ring that are
excluded from the aperture.  Finally, adopting $E(B-V)$=0.16
(Fitzpatrick \& Walborn 1990) and $R$=3.1, we multiply the observed
reverse shock flux by a correction factor of 1.453, to estimate a
dereddened broad H$\alpha$ flux of 1.99($\pm$0.22)$\times$10$^{-13}$
ergs s$^{-1}$ cm$^{-2}$.

\section{MASS FLUX ACROSS THE REVERSE SHOCK}

The total dereddened flux we measure at Earth corresponds to a total
luminosity in the broad H$\alpha$ line of about 5.63$\times$10$^{34}$
ergs s$^{-1}$ at a distance of $\sim$50~kpc, or roughly
15~$L_{\odot}$.  From this we can infer the flux of neutral hydrogen
atoms across the reverse shock.  For each neutral H atom crossing the
reverse shock, roughly 1 Ly$\alpha$ and 0.21 H$\alpha$ photons will be
emitted (Michael et al.\ 2003). Thus, dividing the intrinsic H$\alpha$
luminosity by the energy per H$\alpha$ photon and multiplying by a
factor of 5 gives a total luminosity of hydrogen atoms across the
reverse shock of $\dot{N}_H \simeq 8.9 \times$10$^{46}$ s$^{-1}$, or a
total hydrogen mass flux of roughly 2.3$\times$10$^{-3}$ $M_{\odot}$
yr$^{-1}$.  Michael et al.\ (2003) found that the main emitting
surface area of the reverse shock was within $\pm$30\arcdeg\ of the
equator, just inside the nebular ring.  To first order, the density of
neutral H atoms in the debris prior to crossing the reverse shock is
then

\begin{displaymath}
n_H \simeq \frac{ \dot{N_H} t }{ 4 \pi R_{\rm RS}^3 \tan 30\arcdeg }
\end{displaymath}

\noindent where $\dot{N}_H$=8.9$\times$10$^{46}$ s$^{-1}$ is the
number of hydrogen atoms crossing the reverse shock, $R_{\rm RS}$ is
the present radius of the reverse shock, and $t=$18 yr is the time
since outburst.  Here we have assumed that the supernova debris are in
free expansion, so that $R_{\rm RS}= V_{{\rm H}\alpha} t$.  Taking the
observed velocity $V_{{\rm H}\alpha} \simeq 10^4 $ km s$^{-1}$, we
find $R_{\rm RS}\simeq$0.16 pc.  This is roughly 80\% of the radius of
the forward shock, taken to be the radius of the ring.  This ratio is
close to the theoretically-expected value for self-similar expansion
(Chevalier 1982).  The H$\alpha$ flux we measure implies
$N_H\simeq$60--70 cm$^{-3}$.

Figure 2 shows the history of $\dot{N}_H$ as inferred from STIS
observations of broad H$\alpha$ and Ly$\alpha$ since 1997 (Sonneborn
et al.\ 1998; Michael et al.\ 1998; Heng et al., in prep.) and the
present observation.  The estimated error bars of the values derived
from the STIS observations are large mainly because the corrections
for narrow emission lines were substantially greater in the lower
dispersion STIS spectra.  From Figure 2, we see that $\dot{N}_H$ has
increased by a factor $\sim$4 since the first observation of broad
Ly$\alpha$ from the reverse shock in 1997 (Sonneborn et al.\ 1998).

\section{WILL PREIONIZATION SHUT OFF THE REVERSE SHOCK EMISSION?}

The hot shocked gas lying immediately outside the reverse shock
surface is a luminous source of ionizing photons, roughly half of
which will propagate inward to photoionize hydrogen atoms in the
supernova debris before they reach the reverse shock.  If the
luminosity of these ionizing photons exceeds that of the hydrogen
atoms, the broad Ly$\alpha$ and H$\alpha$ emission will vanish.  Will
this event take place, and if so, when?

We can estimate the luminosity of ionizing photons from the forward
shock of SNR1987A from {\it Chandra} X-ray observations.  Spectral
analysis of the LETG observations (Zhekov et al.\ 2005, also Zhekov et
al.\ in preparation) shows that a two-shock model with temperatures of
0.51 and 2.7~keV, respectively, gives an excellent fit to the X-ray
spectrum at $t = 17.5$ years.  For such a model, most of the ionizing
photons are EUV photons having energies well below the 0.4 -- 10 keV
{\it Chandra} band.  From the two-shock model spectrum, we estimate an
inward luminosity of ionizing ($> 13.6$ eV) photons of $F_i \simeq 3.7
\times 10^{45}$ s$^{-1}$, where we have included a factor of 1.4 to
account for the brightening of X-rays that took place between $t =
17.5$ and $t = 18.0$ years, and a factor 0.5 to account for the fact
that only half of the ionizing photons will propagate inward.  It
follows, then, that the ratio of ionizing photon luminosity to
hydrogen atom luminosity at $t = 18.0$ years is $R_i = F_i/\dot{N}_H
\simeq 0.04$.

We regard this estimate as conservative because we know that the
complex system of shocks in SNR~1987A must have velocities ranging
from $\sim 150$ km s$^{-1}$ (as observed in the line profiles of the
optical hot spots; Pun et al.\ 2002) to $\sim 300 - 1700$ km s$^{-1}$
(as observed in the X-ray line profiles; Zhekov et al.\ 2005).  We
estimated $F_i$ based on the best model fit to the observed X-ray
spectrum which requires shocks of $\sim 600$~and~$\sim1400$~km
s$^{-1}$, respectively.  This may be a significant underestimate,
because shocks having velocities in the range $150 \leq V_S \leq 500$
km s$^{-1}$ might contribute substantially to $F_i$ but relatively
little to the {\it Chandra} band.

At present, the X-ray luminosity of SNR1987A is increasing by a factor
of $\sim$1.7 every year (Park et al. 2005), more rapidly than the
luminosity of hydrogen atoms across the shock.  If present trends
continue, as indicated in Figure~\ref{fig:photo}, $F_i$ should
overtake $\dot{N}_i$ by about 2012 to 2014.  This event could occur
1--2 years later if the expanding SN ejecta have a high He abundance,
because absorption by He atoms could reduce the effective value of
$F_i$.  We regard the prediction in Figure 2 as very conservative,
however.  As we have suggested, we may have underestimated the value
of $F_i$ at 18 years.  Moreover, Luo et al.\ (1994) predicted that,
once the blast wave envelops the circumstellar ring, the ionizing
luminosity should rise very rapidly to a value $F_i \simeq 2 \times
10^{48}$ s$^{-1}$.  This rapid rise, and the consequent vanishing of
the reverse shock emission, could happen anytime during the next
several years.

Additional uncertainty arises from the extrapolation of $\dot{N}_H$.
There are two hydrodynamic scenarios that might bracket its evolution.
The first is the self-similar solution describing the expansion of a
supernova envelope with density law $\rho(v,t) \propto t^{-3}v^{-n}$
into a uniform circumstellar medium (Chevalier 1982).  In this
scenario, the radii of the blast wave and the reverse shock both
increase as $R \propto t^{(n-3)/n}$, and the mass luminosity increases
as $\dot{N}_H \propto t$.  We regard this behavior as a probable lower
limit to the rate of increase of $\dot{N}_H$.  But $\dot{N}_H$ must
begin to increase more rapidly in the near future.  The blast wave is
now overtaking the dense hotspots on the circumstellar ring, and each
such encounter will send a reflected shock inward toward the reverse
shock.  The reflected shocks will eventually merge with the reverse
shock and bring it nearly to a halt.  Thereafter, $\dot{N}_H \propto
t^{(n-4)}$, or $\propto t^5$ for a typical value $n = 9$ (Eastman \&
Kirshner 1989). We regard this as a probable upper limit to the future
evolution of $\dot{N}_H$.  These two limiting behaviors of $\dot{N}_H$
are shown as the dashed curves in Figure~\ref{fig:photo}.

\section{DISCUSSION}

When the suppression of the broad H$\alpha$ emission is seen, it will
provide vital insights into the radiation hydrodynamics of the
developing SNR.  For example, if the suppression occurs early, as we
might expect for reasons noted above, it would provide evidence for
substantial ionizing radiation from shocks too slow to contribute
substantially to the observed X-ray spectrum.  Also, we don't expect
the suppression to take place uniformly around the reverse shock.
X-ray images (Park et al. 2005) certainly show a substantial amount of
structure.  Even if the supernova debris have cylindrical symmetry, we
would expect the broad H$\alpha$ and Ly$\alpha$ emission to vanish
first near the brighter sources of ionizing radiation (i.e. the bright
X-ray knots).  With ground-based telescopes, such variation should be
evident in the evolution of the line profiles.

Yogi Berra said, ``It's tough to make predictions, especially about
the future.''  That statement certainly applies to the future
evolution of SNR 1987A.  Even so, we think that the reverse shock
emission is likely to vanish soon enough that it merits continued
vigilance with both ground- and space-based telescopes.  Yogi Berra
also said, ``You can observe a lot just by watching.''

\acknowledgments \scriptsize

N.S.\ was supported by NASA through grant HF-01166.01A from the Space
Telescope Science Institute, which is operated by the Association of
Universities for Research in Astronomy, Inc., under NASA contract
NAS5-26555.  S.Z.\ and K.H.\ were supported by NASA through Chandra
award G04-5072A to the University of Colorado.


\begin{figure*}
\epsscale{0.6}
\plotone{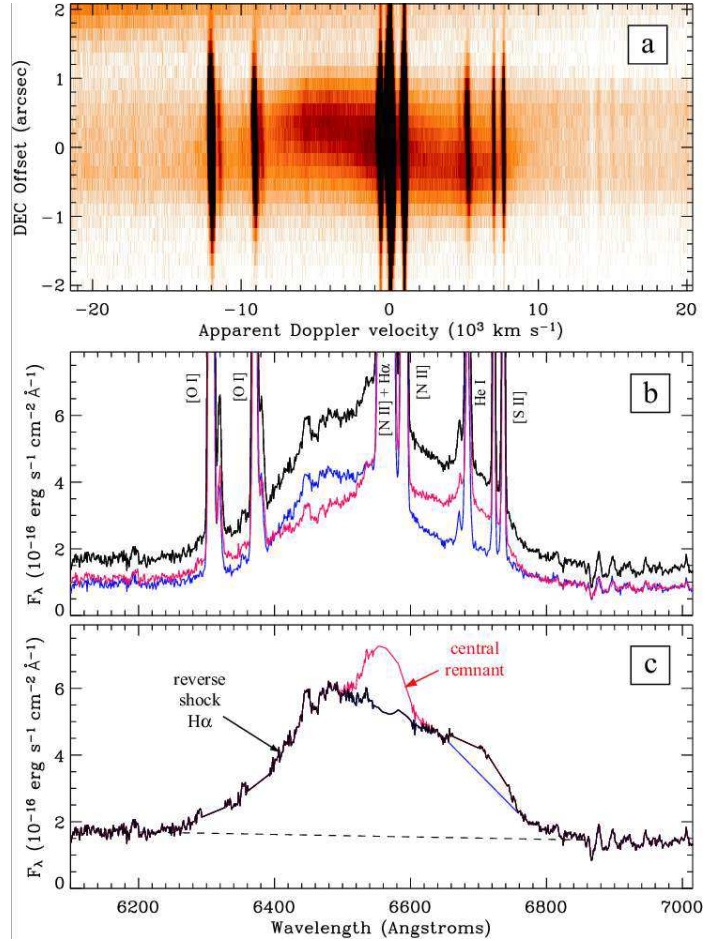}
\caption{LDSS3 spectra of SN1987A showing the H$\alpha$ reverse shock
emission.  (a) Long-slit spectrum showing narrow emission lines from
the ring and hot spots as well as very broad ($\pm$10$^4$ km s$^{-1}$)
H$\alpha$ emission from the reverse shock.  (b) Extracted spectra.
The top tracing (thick black line) is the total emission integrated
over $\pm$1\arcsec\ in the top panel, whereas the thin tracing (blue
in the online edition) is the blueshifted side of the reverse shock at
0--1\arcsec, and the dotted tracing (red in the online edition) is the
redshifted side of the shock at --1--0\arcsec.  (c) Emission from the
reverse shock with the narrow emission components removed.  The thick
black tracing shows the spectrum used to estimate the pure reverse
shock emission.  The dashed tracing (red in the online edition)
includes the H$\alpha$ emission from the central remnant, which would
increase the estimated flux by +6.5\%.  The straight line segment
(blue in the online edition) shows a conservative approximation for
the reverse shock flux interpolated over He~{\sc i} $\lambda$6680 and
the [S~{\sc ii}] doublet, which would be 6\% less than the total flux
in the solid tracing.  The dashed black line shows the continuum level
we chose in measuring the total reverse shock flux.}
\end{figure*}

\begin{figure}
\epsscale{0.6}
\plotone{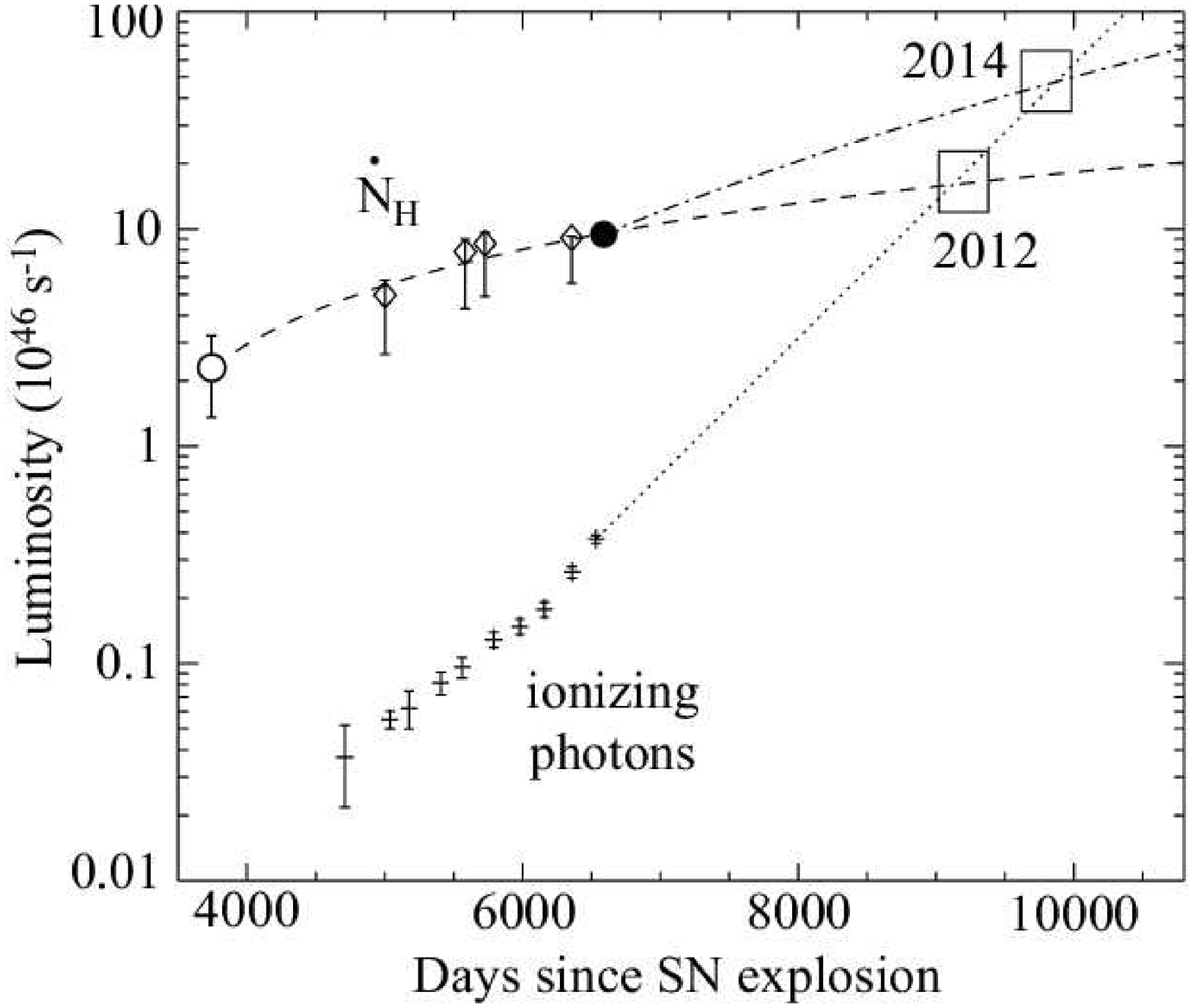}
\caption{The luminosity of H-atoms vs.\ ionizing photons at the
reverse shock in SNR~1987A. The X-ray fluxes used to derive the
ionizing photon curve are taken from Park et al.\ (2005) and then
converted into photon luminosity using the same two-shock model as
explained in the text.  The dotted line represent the increase of a
factor 1.7 yr$^{-1}$, extrapolating from the behavior of the X-ray
data over the past few years.  The H$\alpha$ data from this paper
(solid dot) and from Heng et al.\ (in prep.; diamonds), and Ly$\alpha$
data from Sonneborn et al.\ (1998; unfilled circle) are used to
calculate the H-atom luminosity (see text). The dashed and dash-dotted
lines mark the likely lower limit ($\propto t$) and upper limit
($\propto t^5$) to the future evolution of the H-atom luminosity.}
\label{fig:photo}
\end{figure}

\end{document}